\documentclass[reprint, aip, rsi, amsmath, amssymb, groupedaddress, floatfix]{revtex4-1}
\pdfoutput=1
\usepackage{graphicx}
\usepackage{overpic}

\usepackage{hyperref}

\usepackage{ulem}

\newcommand{\degree}{\ensuremath{^\circ}}
\newcommand{\dee}{\ensuremath{\,\mathrm{d}}}
\newcommand{\pd}[2]{\ensuremath{\frac{\partial#1}{\partial#2}}}
\newcommand{\figref}[1]{Fig.~\ref{#1}}
\newcommand{\apxref}[1]{Appendix \ref{#1}}
\newcommand{\secref}[1]{Sec.~\ref{#1}}
\newcommand{\subsecref}[1]{Subsection~\ref{#1}}
\renewcommand{\exp}[1]{\ensuremath{\mathrm{e}^{#1}}}
\renewcommand{\eqref}[1]{Eq.~(\ref{#1})}

\begin{document}

\title{Velocity map imaging with non-uniform detection: quantitative
  molecular axis alignment measurements via Coulomb explosion imaging}

\author{Jonathan G.~Underwood}
\email{j.underwood@ucl.ac.uk}
\affiliation{Department of Physics and Astronomy, University College
  London, Gower Street, London WC1E 6BT, United Kingdom}

\author{I.~Procino}
\affiliation{Department of Physics and Astronomy, University College
  London, Gower Street, London WC1E 6BT, United Kingdom}

\author{L.~Christiansen}
\affiliation{Department of Chemistry, University of Aarhus, DK-8000
  {\AA}rhus C, Denmark}

\author{J.~Maurer}
\affiliation{Department of Chemistry, University of Aarhus, DK-8000
  {\AA}rhus C, Denmark}

\author{H.~Stapelfeldt}
\affiliation{Department of Chemistry, University of Aarhus, DK-8000
  {\AA}rhus C, Denmark}

\date{\today}

\begin{abstract}
  We present a method for inverting charged particle velocity map
  images which incoorporates a non-uniform detection function. This
  method is applied to the specific case of extracting molecular axis
  alignment from Coulomb explosion imaging probes in which the probe
  itself has a dependence on molecular orientation which often removes
  cylindrical symmetry from the experiment and prevents the use of
  standard inversion techniques for the recovery of the molecular axis
  distribution. By incorporating the known detection function, it is
  possible to remove the angular bias of the Coulomb explosion probe
  process and invert the image to allow quantitative measurement of
  the degree of molecular axis alignment.
\end{abstract}

\maketitle

\section{Introduction}
Photofragment imaging has become a standard tool in the chemical
physicists arsenal of tools for detailed measurements of processes in
gas phase molecules\cite{Whitaker2003, Ashfold2006, Suzuki2006} since
its inception in the 1980s\cite{Chandler1987} and the subsequent
evolution of the technique known as velocity map imaging
(VMI).\cite{Eppink1997}

In a photofragment imaging experiment an expanding sphere of charged
particles is projected onto a two-dimensional (2D) position sensitive
detector. The aim of this detection method is to extract the original
3D distribution of the charged particles from the 2D projection, and
so recover the energy and angular information regarding the
fragmentation process. While it is possible to arrange to image only a
central 2D slice of the 3D distribution using laser or electrostatic
slicing techniques\cite{Townsend2003, Suzuki2006} it is frequently the
case that experimental considerations require imaging of the 2D
projection to regain the 3D distribution.\cite{Bordas1996,
  Winterhalter1999, Whitaker2003, Garcia2004, Dick2014} Reconstructing
the 3D distribution from a single 2D projection requires that the
experiment is cylindrically symmetric about an axis lying in the plane
of the detector; if this condition is not met, then in general it is
only possible to reconstruct the 3D distribution tomographically from
multiple 2D projections.\cite{Smeenk2009, Wollenhaupt2009}

However, there exists another class of experiments in which the
experiment lacks cylindrical symmetry due to a non-uniform detection
function that is otherwise measurable or known. In these circumstances
we show here that it is possible to invert a single 2D image to
recover the 3D distribution while simultaneously correcting for the
non-uniform detection function.

Our methodology is motivated by the desire to characterize the degree
of molecular axis alignment and orientation produced through the
interaction of a molecular sample with intense non-resonant laser
fields.\cite{Larsen2000, Rosca-Pruna2001a, Dooley2003,
  Stapelfeldt2003, Lee2006, Holmegaard2009, Pentlehner2013} For such
aligned/oriented samples to be useful for subsequent experiments, such
as x-ray\cite{Filsinger2011, Kupper2014} or electron
diffraction,\cite{Spence2004, Hensley2012, Miller2014} high-order
harmonic generation,\cite{Itatani2004, Torres2007} and
photodissociation or photoionization studies,\cite{Kumarappan2008,
  Hockett2009, Boll2013} it is necessary to characterize and quantify
the degree of alignment and orientation produced. It is common
practice to utilize Coulomb explosion imaging (CEI) for this
characterization\cite{Larsen2000, Rosca-Pruna2001a, Dooley2003,
  Lee2006, Holmegaard2009, Pentlehner2013} where an intense probe laser
pulse with duration much shorter than molecular rotation is used to
remove multiple electrons from the molecules under study. The multiply
ionized molecules subsequently undergo rapid fragmentation due to
Coulomb repulsion, and imaging the resulting ion fragments is then
used to establish the orientation of the molecules prior to
ionization; under the assumptions that the fragmentation happens
rapidly with respect to rotation, and that the fragments recoil in the
direction of molecular bonds, there is a direct correlation between
the fragment recoil and the molecular orientation. While this
technique has been very succesful at analysing alignment and
orientation of molecular samples it has proved difficult to fully
quantify the degree of alignment/orientation since the ionization
process in CEI has a strong dependence on the molecular orientation
with respect to the ionizing laser polarization. In fact it has become
common practice to characterize the molecular axis alignment and
orientation in such experiments using expectation values calculated
for the resulting VMI image, such as
$\langle\cos^2\theta_\mathrm{2D}\rangle$, where $\theta_\mathrm{2D}$
is the angle in the plane of the detector measured from the axis of
laser polarization. This expectation value includes anisotropy due to
the CEI probing, and is calculated in lieu of a method suitable for
extracting the true molecular axis distribution. Here we show that it
is possible, under certain circumstances, to remove the effect of the
orientation dependence of the CEI probe from the measured molecular
axis distribution, and so extract the true moments of the axis
distribution. This is possible by making an independent measurement of
the CEI orientation dependence using an isotropic gas under the same
conditions as the alignment measurement which can subsequently be
deconvoluted from the CEI images of aligned/oriented samples.

\section{Inversion of photofragment images with non-uniform detection}
\label{sec:general_methodology}
\begin{figure}
  \includegraphics{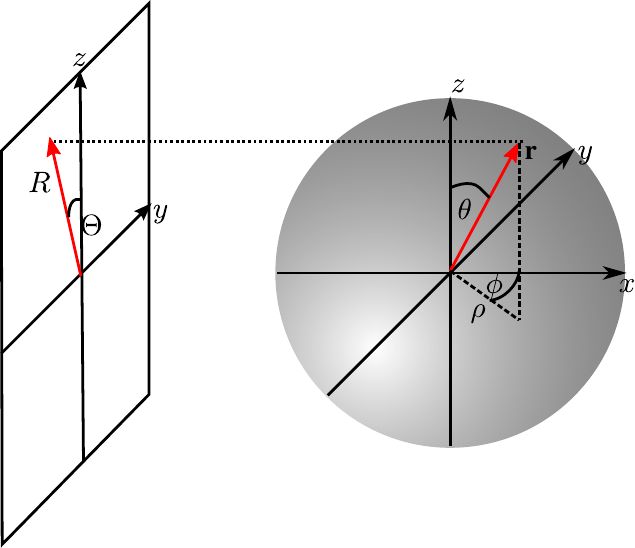}
  \caption{\label{fig:geometry} Relationship between coordinates of
    the original 3D Newton sphere (right) and the 2D projected image
    (left).}
\end{figure}
In a photofragment imaging experiment, the Newton sphere of charged
particles of interest is projected onto a 2D position sensitive
detector via electrostatic lenses which accelerate the charged
particles towards the detector. If the initial distribution is
cylindrically symmetric about an axis parallel to the detector frame,
then the 3D distribution and its 2D projection are related via the
Abel transform,
\begin{equation}
  \label{eq:Abel}
  F(y,z) = 2\int_y^\infty
  \frac{\rho f(\rho,z)}{\sqrt{\rho^2-y^2}}
  \dee\rho,
\end{equation}
where $F(y,z)$ is the 2D projection, $f(\rho,z)$ is the 3D
distribution which is assumed cylindrically about the $z$ axis, and
$\rho$ is the distance to the $z$-axis as illustrated in
\figref{fig:geometry}. Typically, solving the inverse of this equation
to recover the desired $f(\rho,z)$ distribution directly is
sub-optimal due to the sensitivity to experimental noise, and
consequently over the years a number of numerical approaches have been
developed to tackle this problem.\cite{Winterhalter1999, Bordas1996,
  Whitaker2003, Vrakking2001, Garcia2004, Roberts2009}

Here we treat the case where the 3D charged particle distribution
lacks an axis of cylindrical symmetry due to a non-uniform detection
function. We choose to analyse the problem in polar coordinates,
which has been shown to have advantages in terms of localizing any
noise in the inversion process to the very centre of the
image,\cite{Vrakking2001, Garcia2004, Roberts2009} and also provides a
natural description of many optically induced processes in atoms and
molecules.\footnote{However we note that the methodology presented
  here can be applied to other choices of coordinate frames and our
  early work on this resulted in a treatment using cartesian
  coordinates which unfortunately resulted in amplification of
  experimental noise\cite{Procino2011}}

We consider the case where we wish to characterize a 3D cylindrically
symmetric distribution $g(r, \theta)$, which is projected onto the
detector via a non-uniform detection function $D(r, \theta,
\phi)$. Here $r$, $\theta$, and $\phi$ are respectively the radius,
polar angle and azimuthal angle describing the position of a charged
particle on the Newton sphere which is projected onto the detector
(see \figref{fig:geometry}).  We show in \apxref{apx:polar_abel} that
the Abel transformation may be written in spherical polar coordinates
as
\begin{equation}
  \label{eq:FRTheta3}
  \begin{split}
    F(R, \Theta)&=\int_{R}^{\infty}
    \frac{rf(r, \theta, \phi)}{\sqrt{r^2-R^2}}
    \dee r
    \\&=
    \int_{R}^{\infty}
    \frac{rg(r, \theta)D(r, \theta, \phi)}{\sqrt{r^2-R^2}}
    \dee r.
  \end{split}
\end{equation}
where $F(R, \Theta)$ is the projected (image) distribution, $\Theta$
is the polar angle measured in the detection plane with respect to the
$z$ axis, and $R$ is the distance from the image centre (see
~\figref{fig:geometry}).

Since \eqref{eq:FRTheta3} has a similar form to \eqref{eq:Abel} many
of the numerical approaches to inverting \eqref{eq:Abel} could be
adapted to invert \eqref{eq:FRTheta3} to obtain $g(r, \theta)$. Here
we choose to adapt the widely used pBasex approach\cite{Garcia2004}
and expand the desired distribution $g(r,\theta)$ as a product of
basis functions comprising products of Gaussian radial functions and
Legendre polynomials as angular functions,
\begin{equation}
  \label{eq:basisfns}
  g(r,\theta) = 
  \sum_{k=0}^{k_\mathrm{max}}
  \sum_{l=0}^{l_\mathrm{max}}
  C_{kl}g_{kl}(r,\theta),
\end{equation}
where the basis functions are
\begin{equation}
  g_{kl}(r,\theta)=
  \frac{1}{\sigma\sqrt{2\pi}}
  \exp{-\frac{(r-r_k)^2}{2\sigma^2}}
  P_l(\cos\theta).
\end{equation}
Each radial function has a Gaussian width of $\sigma$ and is centred
at $r_k=\Delta r_{k} k$ where $\Delta r_{k} =
r_\mathrm{max}/k_\mathrm{max}$ and $r_\mathrm{max}$ is the maximum
radius of the charged particle cloud considered.

The VMI image $F(R, \Theta)$ can then be written as an expansion in
the corresponding projected basis functions,
\begin{equation}
  \label{eq:image_expansion}
  F(R, \Theta)=
  \sum_{k=0}^{k_\mathrm{max}}
  \sum_{l=0}^{l_\mathrm{max}}
  C_{kl}F_{kl}(R, \Theta),
\end{equation}
where the projected basis functions are given by
\begin{equation}
  F_{kl}(R, \Theta)=
  \int_{R}^{\infty}
  \frac{rg_{kl}(r,\theta)D(r, \theta, \phi)}{\sqrt{r^2-R^2}}
  \dee r.
\end{equation}
In the common case where the projected image is detected on a discrete
grid of cartesian pixels, so long as we choose the width $\sigma$ of
the radial basis functions in \eqref{eq:basisfns} to be around 1
pixel, we can express the image \eqref{eq:image_expansion} in discrete
form as
\begin{equation}
  \label{eq:pbasex_system}
  F_{ij}(R_i, \Theta_j)=
  \sum_{k=0}^{k_\mathrm{max}}
  \sum_{l=0}^{l_\mathrm{max}}
  C_{kl}F_{kl}^{ij}(R_i, \Theta_j),
\end{equation}
where $i$ and $j$ are the detector radial and angular pixel indices
respectively. Here $R_i=(i+\tfrac{1}{2})\Delta R$ is the value of $R$
at the centre of pixel $(i,j)$, where $\Delta R$ is the radial pixel
width, and $\Theta_j=(j+\tfrac{1}{2})\Delta\Theta$ is the value of
$\Theta$ at the centre of the pixel $(i,j)$, where $\Delta\Theta$ is
the angular pixel width. The corresponding discretized basis functions
are given by
\begin{equation}
  \label{eq:projected_basisfns}
  F_{kl}^{ij}(R_i, \Theta_j)=
  \int_{R_i}^{\infty}
  \frac{rg_{kl}(r,\theta)D(r, \theta, \phi)}{\sqrt{r^2-R_i^2}}
  \dee r,
\end{equation}
where 
\begin{align}
  \cos\theta&=
    \frac{R_i\cos\Theta_j}{r}
  \\
  \sin\phi&=
  \frac{R_i\sin\Theta_j}{r\sin\theta}
  =
  \frac{R_i\sin\Theta_j}{\sqrt{r^2-R_i^2\cos^2\Theta_j}}
\end{align}
The basis functions \eqref{eq:projected_basisfns} represent the VMI
images corresponding to each basis function $g_{kl}(r,\theta)$
convoluted with the detection function $D(r, \theta, \phi)$. These
projected basis functions may be calculated using standard numerical
integration methods such as CQUAD in the GSL library\cite{GSL} to
perform the integration over $r$ in \eqref{eq:projected_basisfns}. In
order to obtain the expansion coefficients $C_{kl}$, the system of
linear equations represented by the matrix equation
\eqref{eq:pbasex_system} can then be solved using linear algebra
techniques such as singular value decomposition.\cite{Garcia2004, GSL}
We note that the method as described is identical to
pBasex\cite{Garcia2004} in the limit $D(r, \theta, \phi)=1$.

Once fitted, the $C_{kl}$ coefficients can be used to characterize the
charged particle distribution by calculating the angular integrated
radial spectrum (which is related to the speed distribution of the
particles) according to
\begin{equation}
  \label{eq:radial_spec}
  I(r) = 
  \frac{1}{\sigma\sqrt{2\pi}}
  \sum_{k=0}^{k_\mathrm{max}}
  C_{k0}
  \exp{-\frac{(r-r_k)^2}{2\sigma^2}}.
\end{equation}
In addition, it is usual to characterize the (radially dependent)
angular distribution according to an expansion in Legendre polynomials
$P_l(\cos\theta)$,
\begin{equation}
  \label{eq:angdist}
  f(\theta;r)=
  \frac{1}{\sqrt{4\pi}}
  \sum_{l=0}^{l_\mathrm{max}}
  \beta_l(r)P_l(\cos\theta),
\end{equation}
where the $\beta_l$ coefficients are calculated from the $C_{kl}$
coefficients as
\begin{equation}
  \label{eq:beta_spec}
  \beta_l(r)=
  \frac {
    \sum_{k=0}^{k_\mathrm{max}}
    C_{kl}
    \exp{-\frac{(r-r_k)^2}{2\sigma^2}}
  }
  {
    \sum_{k=0}^{k_\mathrm{max}}
    C_{k0}
    \exp{-\frac{(r-r_k)^2}{2\sigma^2}}
  },
\end{equation}
which are normalized to $\beta_0(r)=1$.

\section{Application to molecular axis alignment and orientation
  probed by Coulomb explosion imaging}
\subsection{Methodology}
\label{sec:methodology}
We turn now to the application of the formalism presented in
\secref{sec:general_methodology} to the measurement of molecular axis
alignment and orientation from Coulomb explosion imaging with VMI
detection.\cite{Larsen2000, Rosca-Pruna2001a, Dooley2003, Lee2006,
  Holmegaard2009, Pentlehner2013}

As described in the introduction, in such experiments, the molecular
sample is first aligned/oriented with strong non-resonant laser fields
(and sometimes static electric fields). Subsequently, in order to
measure the degree of alignment/orientation produced, an intense probe
laser pulse with duration much shorter than molecular rotation is used
to remove multiple electrons from the molecules under study. The
multiply ionized molecules subsequently undergo rapid fragmentation
due to Coulomb repulsion, and imaging of the resulting ion fragments is
then used to establish the molecular axis distribution prior to
ionization. Under the assumptions that the fragmentation happens
rapidly with respect to rotation, and that the fragments recoil in the
direction of molecular bonds, there is a direct correlation between
the fragment recoil and the molecular orientation. As mentioned
previously, the challenge here is to deconvolute the non-uniform
orientational response of the Coulomb explosion process from the
measurement in order to yield the molecular axis distribution prior to
Coulomb explosion. The strategy we propose here is:
\begin{enumerate}
\item Perform a CEI measurement on an isotropic gas sample, with the
  CEI probe polarization direction chosen such that an axis of
  cylindrical symmetry is contained in the plane of the detector.
\item Invert the image from step 1 above to obtain the 3D
  distribution of CEI ions by solving \eqref{eq:pbasex_system} with
  $D(r, \theta, \phi)=1$. Since this distribution was obtained with an
  isotropic gas sample, we can obtain the orientational dependence of
  the CEI probe process for the CEI probe laser polarization state,
  pulse duration and intensity employed in step 1 above.
\item Perform a CEI measurement on the aligned/oriented molecular sample
  using the same probe polarization state, pulse duration and
  intensity as used in step 1 above.
\item Invert the VMI image from step 3 using
  \eqref{eq:pbasex_system} and a detection function derived from step
  (2) in order to deconvolute the orientational dependence of the CE
  process from the observed fragment distribution, and so obtain the
  molecular axis distribution.
\end{enumerate}
We note that we require the molecular axis distribution in step 3
above to have an axis of cylindrical symmetry lying in the plane of
the detector in order to apply \eqref{eq:pbasex_system}. However, step
3 does \emph{not} require that the same geometry of the probe is used
as for step 2. For example, if using a linearly polarized probe, step
2 requires that the linear polarization lies in the plane of the
detector, but in step 3 we are free to rotate the probe polarization
to lie in a different direction such as perpendicular to the detector
plane. For this reason we introduce two frames of reference: (i) the
detection frame (DF); and (ii) the axis distribution frame (AF). In
order to invert the observed image in step 2, we require that the DF
possesses an axis of cylindrical symmetry lying in the plane of the
detector in step 1. As such, the detection function will have an axis
of cylindrical symmetry in the DF. We also require that the molecular
axis distribution possesses an axis of cylindrical symmetry in the
plane of the detector in step 3. However, in step 3, the DF may be
chosen to lie in any direction relative to the AF.

The inversion of the image recorded for the isotropic molecular sample
in step 2 will yield the fit coefficients for the distribution of
fragments, $C'_{kl}$. Here, and in what follows, we use a prime to
denote properties relating to the detection function. Under the
assumption that the fragments recoil along the direction of the
molecular bond, this distribution will correspond to the probability
of CE for each orientation of that bond relative to the laser
polarization, and so these coefficients can be used to construct the
orientational detection function required for step 4.

In the reference frame defined by the detection laser polarization, we
can write the orientational dependence of the CE probe as
\begin{equation}
  \label{eq:detfn_DF}
  D(\theta'; r) = \frac{1}{\sqrt{4\pi}}
  \sum_l^{l_\mathrm{max}}
  \beta'_l(r)
  P_l(\cos\theta'),
\end{equation}
where $\theta'$ is measured relative to the cylindrical symmetry axis
in the DF. The expansion coefficients $\beta'_l(r)$ are calculated from
the fit coefficients $C'_{kl}$ according to \eqref{eq:beta_spec}.

In order to construct the basis functions
\eqref{eq:projected_basisfns} for step 4, we need to calculate the
detection function in the AF. The angular dependence of the detection
function expressed in the DF, $D(\theta', \phi';r)$ is related to the
angular dependence of the detection function in the AF, $D(\theta,
\phi;r)$ through a rotation through the Euler angles $(\alpha, \Omega,
\gamma)$ connecting the AF and DF.~\cite{Zare}  For the present case
where the detection function has an axis of cylindrical symmetry, we
can set $\gamma=0\degree$. In \apxref{apx:detfn_rotn} we show that the
detection function in the AF may be expressed in terms of the
$\beta'_l(r)$ coefficients found from the probe-alone data inversion
as
\begin{equation}
  \label{eq:detfn_af}
  D(\theta, \phi; r)= \frac{1}{\sqrt{4\pi}}
  \sum_l
  \beta'(r)
  P_l(\cos\Delta),
\end{equation}
where
\begin{equation}
  \label{eq:detfn_loc}
  \begin{split}
    \cos\Delta&=
    \cos\Omega\cos\theta +
    \sin\Omega\sin\theta\cos(\alpha-\phi)
    \\
    &=\cos\Omega\cos\theta +
    \sin\Omega\sin\theta
    (\cos\alpha\cos\phi+\sin\alpha\sin\phi)
    .
    \end{split}
\end{equation}
Eqs.~(\ref{eq:detfn_af}) and (\ref{eq:detfn_loc}) allow for the
evaluation of $D(r, \theta, \phi)$ in \eqref{eq:projected_basisfns}
during the numerical integration over $r$ when calculating the basis
functions.

It is important to note that steps 1 and 2 allow us to retrieve
$D(\theta,\phi;r)$, a detection function dependent upon two of the
Euler angles $(\theta,\phi)$ describing molecular orientation in the AF. As such, this detection function is averaged over the third Euler angle $\chi$ that would be needed to specify the molecular orientation.~\cite{Zare} This angle describes rotation of the molecule around the molecular
$z$-axis. As such, this strategy is applicable to extracting the
alignment/orientation of linear molecules and symmetric rotor
molecules. For asymmetric rotor molecules where localization in $\chi$
accompanies localization in $\theta$~\cite{Larsen2000, Underwood2005,
  Lee2006} care must be taken, and this approach will only apply when
either the localization in $\chi$ is small and/or
$D(\theta, \phi, \chi; r)$ is independent of $\chi$. The latter
situation arises for many molecules.

\subsection{Experimental example}

\begin{figure*}[t]
  \centering
  \includegraphics{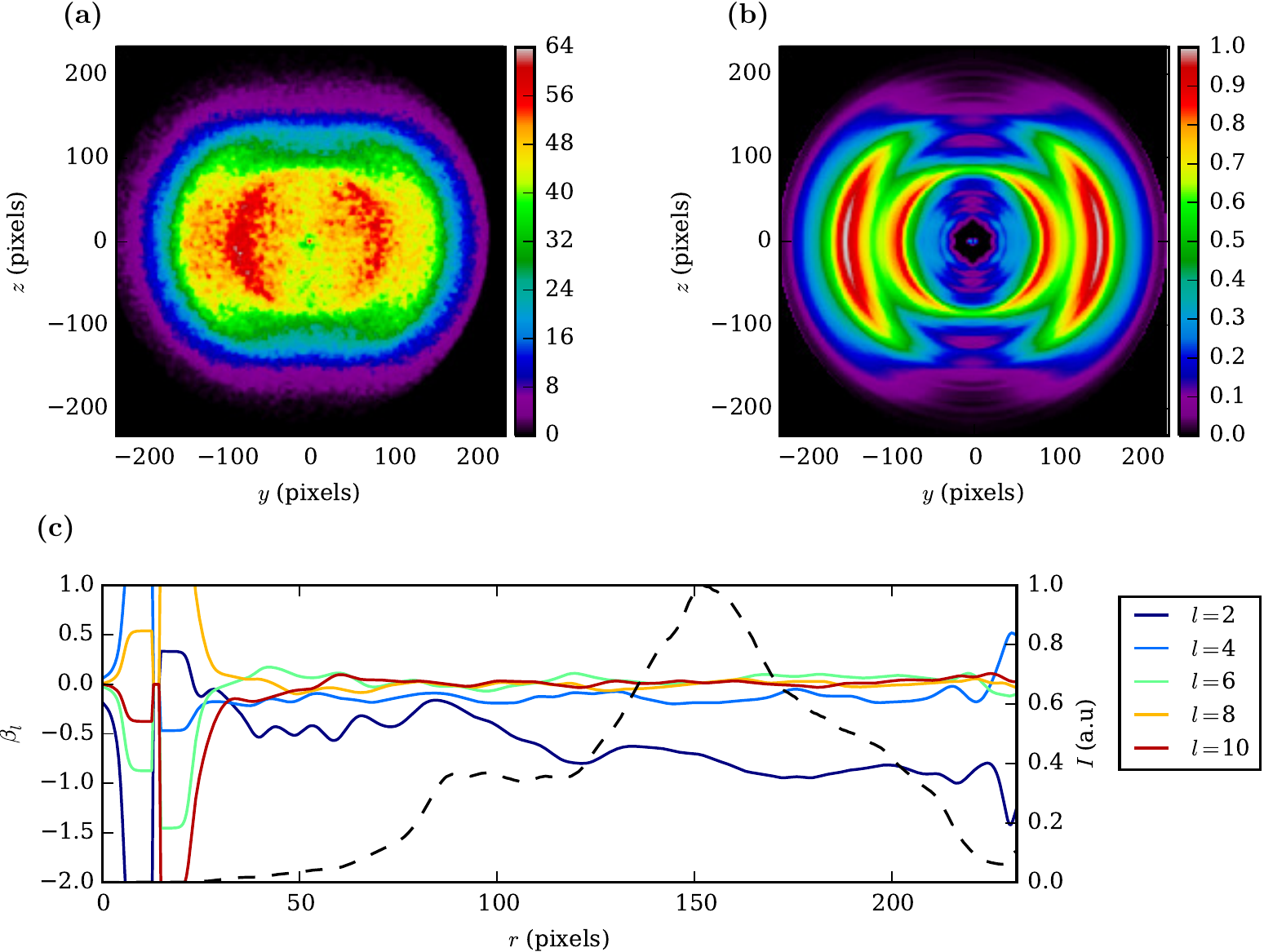}
  \caption{(a) Experimental I$^+$ VMI image recorded for the
    circularly polarized probe laser alone. The axis of cylindrical
    symmetry (corresponding to the laser propagation direction) lies
    parallel to the $z$-axis. (b) Corresponding pBasex inverted I$^+$
    image. (c) Radial dependence of the $\beta_l(r)$ angular
    parameters corresponding to the pBasex inverted image (solid
    lines). The radial spectrum is also shown (dashed line).}
  \label{fig:probe}
\end{figure*}

As a demonstration of the approach outlined in
\subsecref{sec:methodology}, we apply the strategy to the retrieval of
the molecular axis distribution of a sample of 1,4-diiodobenzene
(pDIB) molecules aligned with a strong laser field at 1064~nm
wavelength.\cite{Kumarappan2006} For this molecular species, a strong
linearly polarized laser field will induce alignment of the I-I axis
(the most polarizable axis) towards the laser field polarization
direction. CEI was used to characterize the alignment of the I-I axis,
through detection of recoiling I$^+$ fragments. Under the assumption
that the I$^+$ fragments recoil axially along the C-I bonds in the
molecule, the I$^+$ recoil direction maps directly to the I-I axis
direction in the lab frame. For this molecule, we expect the
dependence of the detection function on the angle $\chi$ to be
negligible.

The molecular sample was prepared in a molecular beam with a
rotational temperature of ca.~1-2~K. The linearly polarized alignment
laser field had a pulse duration of 10~ns which is much longer than
the time scale for molecular rotation. Consequently, this laser field
adiabatically induces molecular axis alignment of the I-I axis in the
sample, with maximal alignment occurring at the peak of the laser
field.~\cite{Friedrich1995, Stapelfeldt2003} Subsequently a second
probe laser pulse at 800~nm and with duration of 30~fs was timed to
arrive at the peak of the alignment laser field. This high intensity
laser pulse induced Coulomb explosion of the aligned molecules, and
the I$^+$ ions produced were detected with a VMI spectrometer.

We report here the results of two different studies. In the first
study a circularly polarized probe laser pulse was employed, and the
alignment using two different intensities of the aligning laser pulse
are compared. In the second study a linearly polarized probe laser
pulse was employed, and we examine the effect of the probe geometry
employed by comparing images recorded with the probe polarization
either parallel or perpendicular to the aligning laser polarization.

\subsubsection{Circularly polarized probe pulse}
\label{sec:circ_probe}
Here we detail experiments carried out with a circularly polarized
laser pulses with intensity of $2\times 10^{14}$~W/cm$^2$. In
\figref{fig:probe}a we show the I$^+$ VMI data recorded for the
circularly polarized probe alone. Two radially separated rings are
seen corresponding to two different CE channels.  The outermost
channel corresponds to CE of triply charged pDIB molecules whereas the
innermost ring results from CE of doubly charged pDIB
molecules.~\cite{ChristensenPre} \figref{fig:probe}b shows the
distribution of I$^+$ ions obtained from the pBasex inversion of the
experimental VMI image. This inversion was carried out by binning the
experimental image into a 256$\times$256 polar image and solving
\eqref{eq:pbasex_system} with $k_\mathrm{max}=128$,
$l_\mathrm{max}=10$, $D(r, \theta, \phi)=1$ (i.e.~uniform detection),
and $\sigma=1.2$ pixels. The coefficients $C_{kl}$ in
\eqref{eq:pbasex_system} were obtained through projected Landweber
iteration~\cite{Bertero1998, Renth2006} with a projection function
setting $C_{kl}=0$ if $C_{k0} <0$ at each iteration. Further, due to
the inversion symmetry of the experiment, only even values of $l$ were
included in \eqref{eq:pbasex_system}.

In \figref{fig:probe}c we show the $\beta_l(r)$ parameters calculated
according to \eqref{eq:beta_spec}, as well as the radial spectrum
calculated from \eqref{eq:radial_spec}. From this plot we can see that
in regions with non-negligible ion intensity, contributions from
$\beta_l(r)$ parameters with $l>4$ are negligible, and as such the
probe detection function is well defined by $C_{kl}$ coefficients with
$l\le 4$.

The distribution shown in \figref{fig:probe}b represents the detection
function for CEI probing with the circularly polarized laser pulse for
the intensity and pulse duration used. It is this distribution that
samples the aligned axis distribution in the subsequent measurements
with laser-aligned molecular samples.

\begin{figure*}
  \centering
  \includegraphics{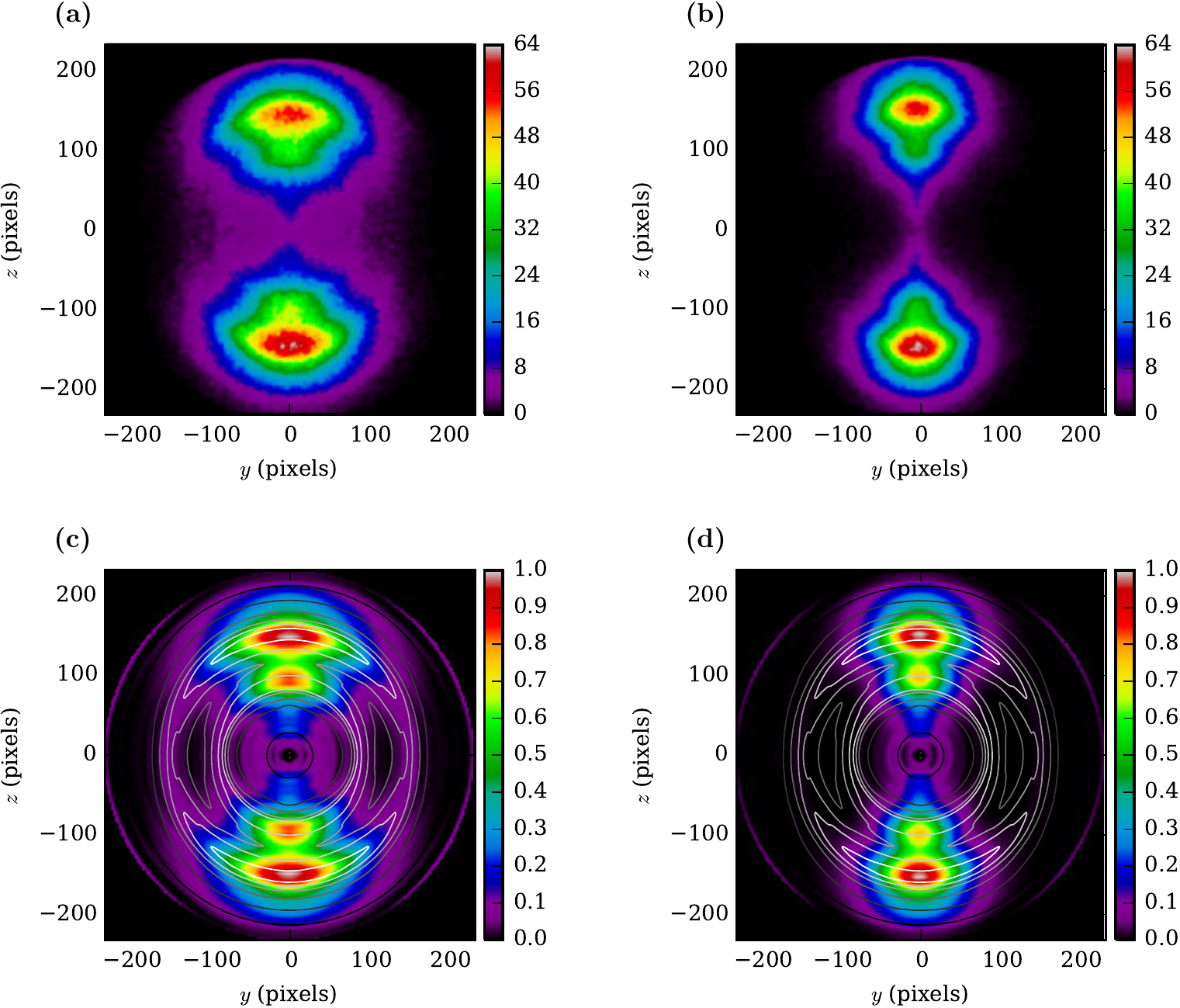}
  \caption{ (a) and (b): Experimental I$^+$ VMI images recorded for
    laser aligned pDIB probed via Coulomb explosion with a circularly
    polarized laser pulse. The aligning laser polarization is along
    $z$, and the probe propagation direction lies along $y$. Images
    are shown for aligning laser field intensities of (a) $1.5\times
    10^{11}$~W/cm$^2$, and (b) $7.7\times 10^{11}$~W/cm$^2$.  (c) and
    (d): Corresponding pBasex inverted I$^+$ images with detection
    function deconvoluted. Overlaid is a grayscale contour map
    corresponding to the probe alone distribution shown in
    \figref{fig:probe}b.}
   \label{fig:aligned_vmi}
\end{figure*}

\begin{figure}
  \includegraphics{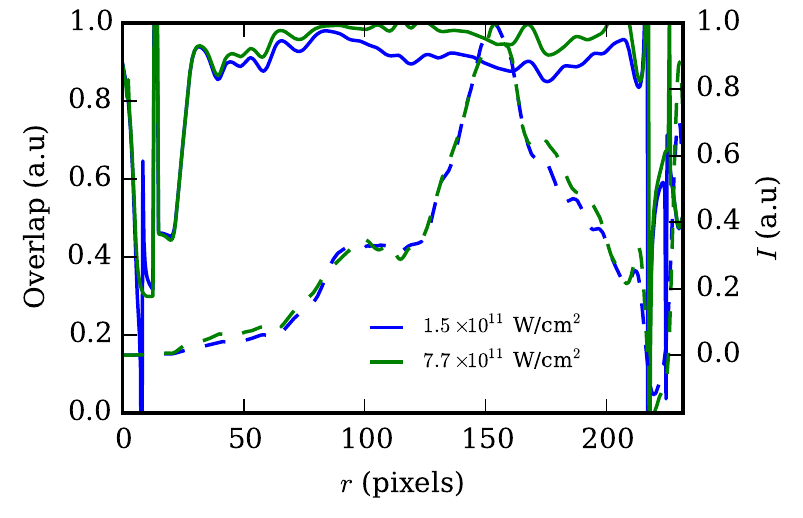}
  \caption{Radial dependence of the angular overlap factor
    corresponding to the pBasex inversion of the images shown in
    \figref{fig:aligned_vmi} (solid lines) for alignment laser
    intensities of $1.5\times 10^{11}$~W/cm$^2$ and $7.7\times
    10^{11}$~W/cm$^2$, and a circularly polarized probe. The
    corresponding radial spectra are also shown (dashed lines).}
  \label{fig:DIB_overlap}
\end{figure}

\begin{figure*}
  \centering
  \includegraphics{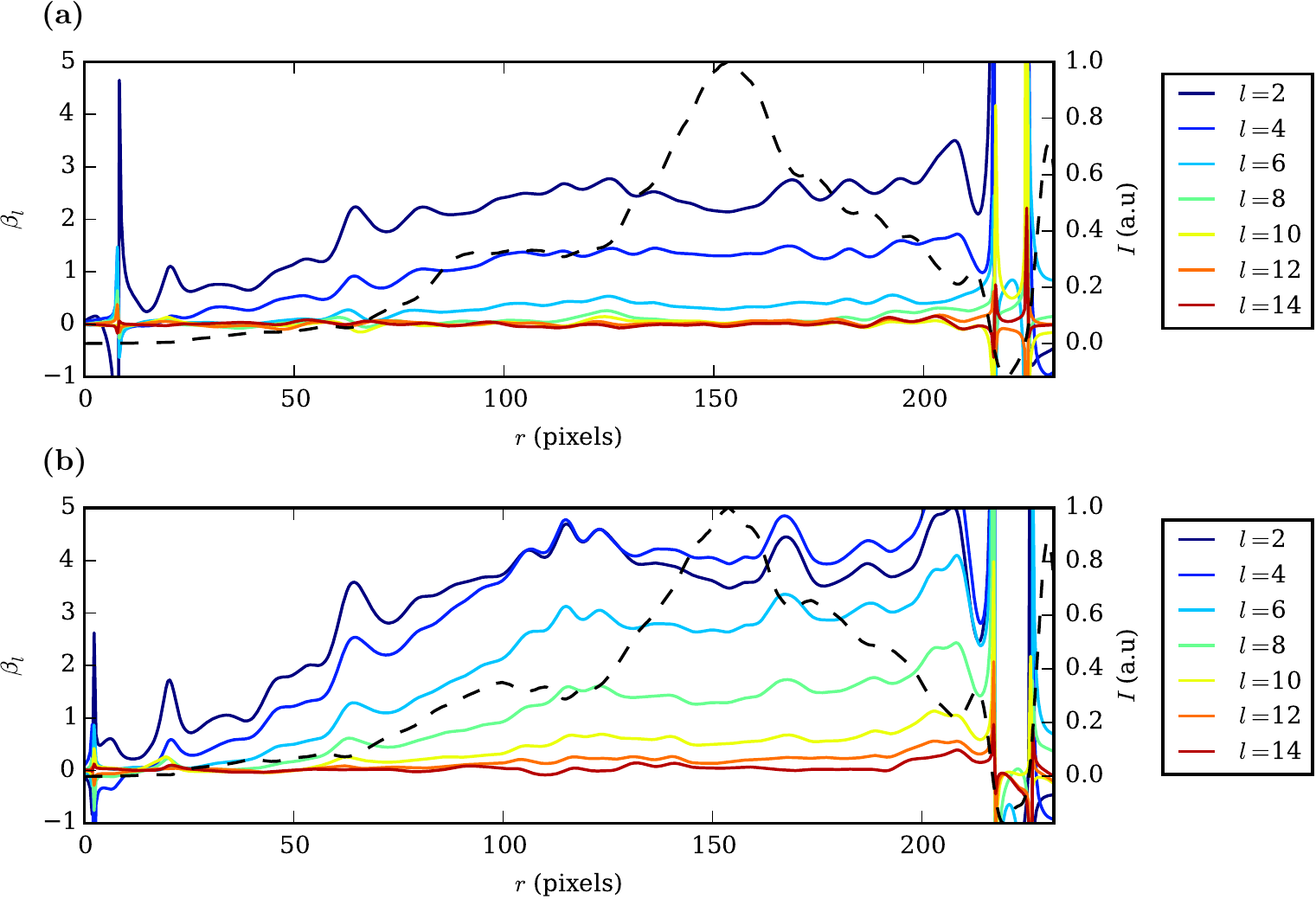}
  \caption{Radial dependence of the $\beta_l(r)$ angular parameters
    obtained from the deconvoluted pBasex inversion of the images
    shown in \figref{fig:aligned_vmi} (solid lines). (a) alignment
    laser intensity of $1.5\times 10^{11}$~W/cm$^2$. (b) alignment
    laser intensity of $7.7\times 10^{11}$~W/cm$^2$. The corresponding
    radial spectra are also shown (dashed lines).}
  \label{fig:DIB_beta}
\end{figure*}

\begin{figure}
  \includegraphics{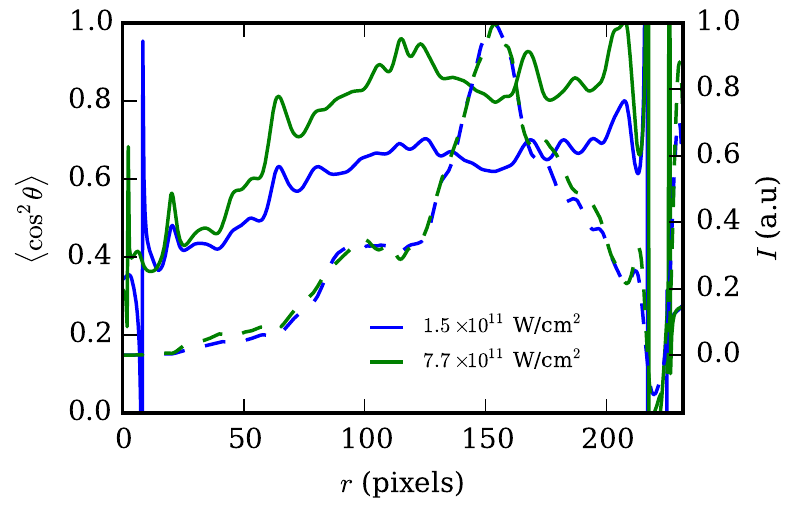}
  \caption{Radial dependence of the $\langle\cos^2\theta\rangle$
    expecatation values resulting from inversion of the images shown
    in \figref{fig:aligned_vmi} (solid lines) for the two alignment
    laser intensities of $1.5\times 10^{11}$~W/cm$^2$ and $7.7\times
    10^{11}$~W/cm$^2$, and a circularly polarized probe laser. The
    corresponding radial spectra are also shown (dashed lines).}
  \label{fig:DIB_cos2}
\end{figure}

In \figref{fig:aligned_vmi}(a) and \figref{fig:aligned_vmi}(b) we
show the I$^+$ VMI data recorded for a sample of molecules aligned
with linearly polarized laser fields of intensities $1.5\times
10^{11}$~W/cm$^2$ and $7.7\times 10^{11}$~W/cm$^2$ respectively. The
aligning laser field was polarized parallel to the $z$-axis. This
laser field therefore induces alignment of the molecular I-I-axis
towards the $z$-axis. The circularly polarized CE pulse propagated
parallel to the $y$-axis such that the light was polarized in the
$xz$-plane. The observed VMI image therefore corresponds to the
molecular axis distribution sampled by the detection function of the
probe.

In \figref{fig:aligned_vmi}c and \figref{fig:aligned_vmi}d we show the
recovered distributions of I$^+$ ions following deconvolution of the
detection function determined from the probe alone data following the
procedure outlined in \subsecref{sec:methodology}.  The inversions to
recover these distributions were carried out with $l_\mathrm{max}=14$
and $k_\mathrm{max}$, $\sigma$ and the number of polar bins the same
as for the probe-alone data. The detection function
$D(r, \theta, \phi)$ used to construct the basis functions
(\eqref{eq:projected_basisfns}) was calculated from
\eqref{eq:detfn_DF} with $\Delta$ calculated from \eqref{eq:detfn_loc}
setting $\Omega=90\degree$ and $\alpha=0\degree$. The coefficients
$C_{kl}$ in \eqref{eq:pbasex_system} were obtained through Landweber
iteration with no projection function.~\cite{Bertero1998, Renth2006}
Further, due to the inversion symmetry of the experiment, only even
values of $l$ were included in \eqref{eq:pbasex_system}.

Overlaid on each recovered distribution in \figref{fig:aligned_vmi}c and
\figref{fig:aligned_vmi}d is a grayscale contour map corresponding to the
probe detection function axially integrated over the azimuthal angle
$\phi$. The calculation of this axially integrated detection function
is detailed in \apxref{apx:detfnint}. This contour map provides a
visual representation of the detection function -- its value
represents the detection probability at each value of $\theta$
integrated over all values of $\phi$. As can be seen from inspection
of \figref{fig:aligned_vmi}c and \figref{fig:aligned_vmi}d, the probe
detection function samples the molecular axis distribution with high
efficiency for both aligning laser intensities.

The degree of overlap of the detection function with the axis
distribution determines the extent to which the full molecular axis
distribution is sampled, and the reliability of the deconvolution
process. We can quantify the degree of this overlap by evaluating the
angular overlap factor
\begin{equation}
  \label{eq:overlap}
  O(r)=\frac{1}{D_\mathrm{max}(r)}
  \int^{2\pi}_0\dee\phi
  \int^{\pi}_0\sin\theta\dee\theta
  f(\theta;r)D(\theta,\phi;r)
  ,
\end{equation}
where $D_\mathrm{max}(r)$ is the maximum value of the angular
dependence of the detection function in the AF,
$D(\theta,\phi;r)$. This integral will take values between 0 (no
overlap) and 1 (maximal overlap). The evaluation of this integral is
detailed in \apxref{apx:overlap}. In \figref{fig:DIB_overlap} we show
the radial dependence of the this overlap factor for the two aligning
laser intensities employed. The overlap factor is clearly lower for
the less well aligned distribution at the lower aligning laser
intensity, reflecting the fact that the broader axis distribution
extends further into the region of lower probability of CE by the
probe laser pulse, as is also seen by comparing \figref{fig:aligned_vmi}c
and \figref{fig:aligned_vmi}d. Nonetheless in both cases the overlap
factor is above 0.9 signifying good sampling of the axis distribution.

In \figref{fig:DIB_beta} we show the $\beta_l(r)$ parameters
calculated according to \eqref{eq:beta_spec}, as well as the radial
spectrum calculated from \eqref{eq:radial_spec}. For both intensities
employed, the resulting $\beta_{14}(r)$ coefficient remained at 0, and
increasing the value of $l_\mathrm{max}$ beyond 14 led to no
significant change in the inverted image. These observations indicate
that the alignment distribution is well characterized with
$l_\mathrm{max}=14$. At the lower intensity of
$1.5\times 10^{11}$~W/cm$^2$ all $\beta_l$ coefficients are seen to be
smaller in magnitude than for the higher intensity of
$7.7\times 10^{11}$~W/cm$^2$, and indeed at the lower alignment laser
intensity the $\beta_{10}(r)$ coefficient was seen to be
negligible. This is consistent with the high alignment laser intensity
producing a higher degree of molecular axis alignment. In both cases,
some large fluctuation in $\beta_l(r)$ values is observed at the
largest values of $r$ due to the experimental image being slightly
truncated by the detector edge.

In \figref{fig:DIB_cos2} we show the $\langle\cos^2\theta\rangle(r)$
expectation values for the data as well as the radial spectrum
calculated from \eqref{eq:radial_spec}. This expectation value is a
commonly used figure-of-merit for characterizing the degree of
molecular axis alignment. It is important to note that this is an
expectation value of the molecular axis distribution, rather than the
commonly used value $\langle\cos^2\Theta\rangle$, referred to
as $\langle\cos^2\theta_\mathrm{2D}\rangle$, which is an expectation
value of the projected image of the axis distribution and which
includes the effect of the non-uniform detection function. The value
of $\langle\cos^2\theta\rangle(r)$ was calculated acording to
\begin{equation}
  \label{eq:cos2_expval}
  \langle\cos^2\theta\rangle(r)=
  \int_{0}^{\pi}
  \sum_{l=0}^{l_\mathrm{max}}
  \beta_l(r)P_l(\cos\theta)
  \cos^2\theta\sin\theta\dee\theta.
\end{equation}

We note that for both aligning laser intensities, the outermost
channel ($r\sim 150$ pixels) indicates a slightly higher degree of
molecular axis alignment than the innermost channel ($r\sim 100$
pixels) -- this is evidenced by the smaller values of the $\beta_l$
parameters for the inner channel compared to the outer channel in
\figref{fig:DIB_beta}, and to a lesser extent by the
$\langle\cos^2\theta\rangle$ expectation values for each channel. As
mentioned, the I$^+$ signal in the outermost channel mainly originates
from Coulomb explosion of triply charged molecules whereas I$^+$ ions
in the innermost channel mainly originate from doubly ionized
molecules. The triply ionized molecules are produced in the region of
the probe laser focus where the intensity is highest and therefore
also in the region where the alignment laser intensity is highest. As
such the outermost channel probes molecules that are expected to be
slightly better aligned than molecules probed by the innermost
channel. Additionally, in the preceding development of our methodology
we have implicitly assumed that the fragment I$^+$ ions recoil axially
along the direction of the molecular I-I axis such that there is a
direct correspondence between fragment recoil and molecular axis
alignment. It is possible that this axial recoil condition of the
I$^+$ fragments is better fulfilled for the outermost channel than for
the innermost channel.  In general, some deviation from axial recoil
is expected due to bonding in the multiply charged molecular ion
created by the probe pulse and to possible charge-asymmetry in the
Coulomb explosion process. The effect of the non-axial recoil is
expected to slightly reduce the degree of measured alignment and it is
not removed by our deconvolution of the detection function -- this
will be discussed further in a future
publication~\cite{ChristensenPre}.

\subsubsection{Linearly polarized probe pulse}
\begin{figure*}[t]
  \centering
  \includegraphics{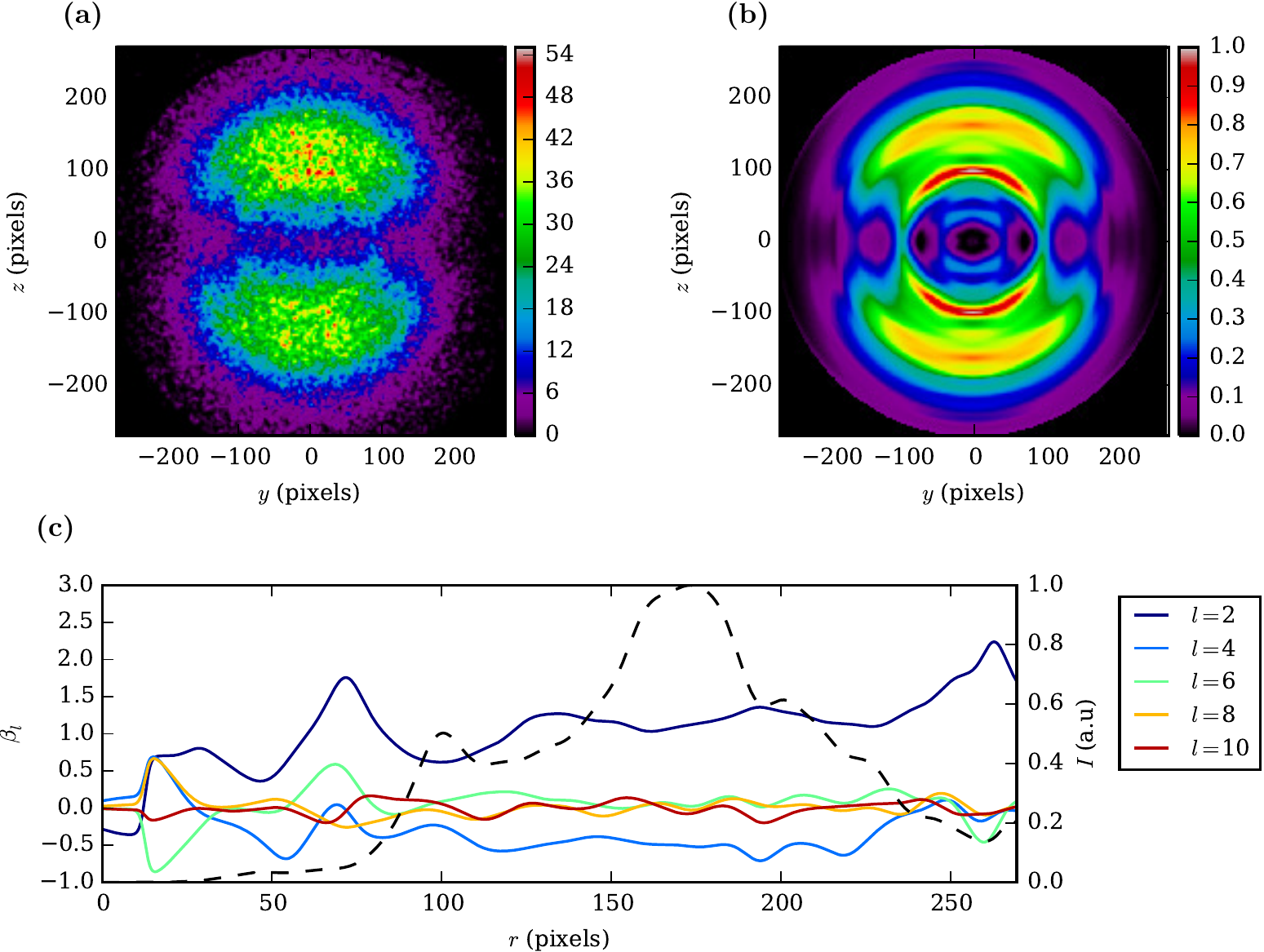}
  \caption{(a) Experimental I$^+$ VMI image recorded for the linearly
    polarized probe laser alone. The axis of cylindrical symmetry
    (corresponding to the laser propagation direction) lies parallel
    to the $z$-axis. (b) Corresponding pBasex inverted I$^+$
    image. (c) Radial dependence of the $\beta_l(r)$ angular
    parameters corresponding to the pBasex inverted image (solid
    lines). The radial spectrum is also shown (dashed line).}
  \label{fig:probe2}
\end{figure*}

Here we detail a second study carried out with a linearly polarized
probe pulse of intensity $3.2\times10^{14}$~W/cm$^2$, and a linearly
polarized aligning pulse of intensity $6.6\times10^{11}$~W/cm$^2$. In
\figref{fig:probe2}a we show the VMI data recorded for the probe laser
alone with its polarization along the $z$-axis. The same two CEI
channels as observed with the circularly polarized probe pulse are
evident in the VMI data. \figref{fig:probe2}b shows the corresponding
distribution of I$^+$ ions obtained from the pBasex inversion of the
VMI image. The lower number of counts in this image (due to a reduced
data collection time) required a more coarse binning of the data when
carrying out the pBasex inversion compared to the circularly polarized
probe case. The experimental image was binned into a 128$\times$128
polar image and \eqref{eq:pbasex_system} was solved with
$k_\mathrm{max}=64$, $l_\mathrm{max}=10$, $D(r, \theta, \phi)=1$
(i.e.~uniform detection), and $\sigma=1.75$ pixels. The coefficients
$C_{kl}$ in \eqref{eq:pbasex_system} were obtained through projected
Landweber iteration~\cite{Bertero1998, Renth2006} with a projection
function setting $C_{kl}=0$ if $C_{k0} <0$ at each iteration. As
previously, only even values of $l$ were included in
\eqref{eq:pbasex_system}.

In \figref{fig:probe2}c we show the $\beta_l(r)$ parameters calculated
according to \eqref{eq:beta_spec}, as well as the radial spectrum
calculated from \eqref{eq:radial_spec}. As was the case with the
circularly polarized probe pulse, in regions with non-negligible ion
intensity, contributions from $\beta_l(r)$ parameters with $l>4$ are
negligible.

\begin{figure*}
  \centering
  \includegraphics{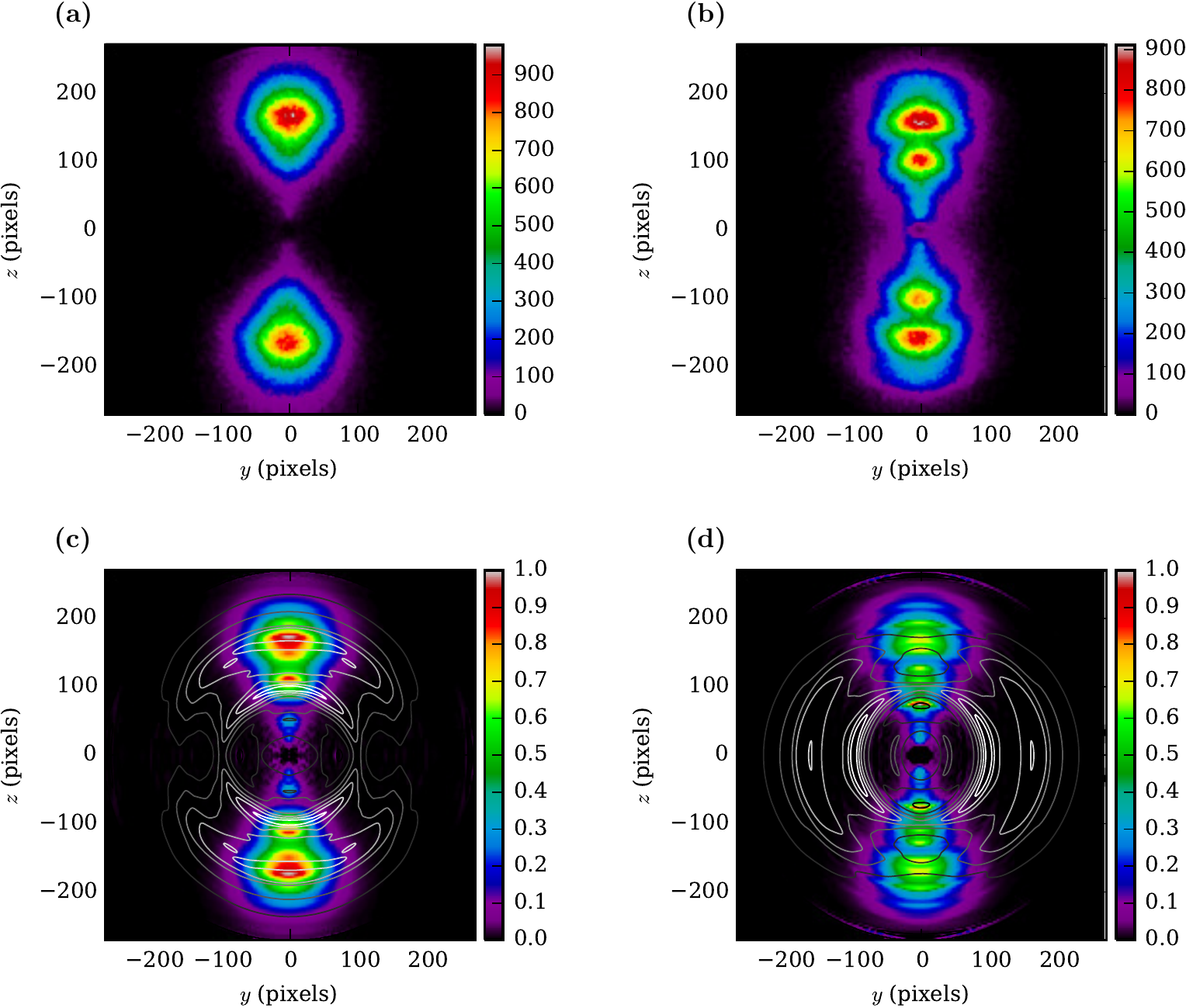}
  \caption{(a) and (b): Experimental I$^+$ VMI images recorded for
    laser aligned pDIB probed via Coulomb explosion with a linearly
    polarized probe laser pulse. The aligning laser pulse was
    polarized along the $z$ direction. In (a) the probe laser was also
    polarized along the $z$-axis. In (b) the probe laser was polarized
    along the $x$-axis (perpendicular to the detector plane). (c) and
    (d): Corresponding pBasex inverted I$^+$ images with detection
    function deconvoluted. Overlaid is a grayscale contour map
    corresponding to the $\phi$-integrated detection function.}
  \label{fig:aligned_vmi_2}
\end{figure*}

In \figref{fig:aligned_vmi_2}a and \figref{fig:aligned_vmi_2}b we show
VMI data recorded for molecules aligned with a linearly polarized laser
field and probed with a linearly polarized probe pulse in two
different geometries. For the data in \figref{fig:aligned_vmi_2}a a
parallel geometry was employed with both the aligning and probe laser
polarizations along the $z$-axis. In \figref{fig:aligned_vmi_2}b a
perpendicular geometry was used with the aligning laser field
polarized along the $z$-axis and the probe laser field polarized along
the $x$-axis (perpendicular to the detection plane). From these images
it is apparent that the relative magnitude of the two CEI channels
depends strongly on the orientation of the probe pulse polarization
relative to the molecular axis.

In \figref{fig:aligned_vmi_2}c and \figref{fig:aligned_vmi_2}d we show
the distributions of I$^+$ ions following deconvolution of the
detection function determined from the probe alone data following the
procedure outlined in \secref{sec:methodology}. Overlaid on each
recovered distribution in \figref{fig:aligned_vmi_2}c and
\figref{fig:aligned_vmi_2}d is a grayscale contour map corresponding
to the probe detection function axially integrated over the azimuthal
angle $\phi$ (\apxref{apx:detfnint}). As can be seen from comparing
\figref{fig:aligned_vmi_2}c and \figref{fig:aligned_vmi_2}d, the probe
detection function samples the axis distribution with much lower
efficiency when the probe polarization is along the $x$-axis,
perpendicular to the direction of molecular alignment. Note that in
\figref{fig:aligned_vmi_2}d, the axially integrated detection function
has a higher efficiency along the $y$-axis than the $z$-axis due to
the integration over $\phi$ encompassing the $x$-axis for
$\theta=90^\circ$.

The inversions to recover the distributions in
\figref{fig:aligned_vmi_2} were carried out by binning the
experimental image into a 256$\times$256 polar image and solving
\eqref{eq:pbasex_system} with $k_\mathrm{max}=128$,
$l_\mathrm{max}=20$, and $\sigma=1.2$ pixels. The detection function
$D(r, \theta, \phi)$ used to construct the basis functions
(\eqref{eq:projected_basisfns}) was calculated from
\eqref{eq:detfn_DF} with $\Delta$ calculated from \eqref{eq:detfn_loc}
setting $\Omega=0\degree$ and $\alpha=0\degree$ for the parallel
geometry and $\Omega=90\degree$ and $\alpha=0\degree$ for the
perpendicular geometry. The coefficients $C_{kl}$ in
\eqref{eq:pbasex_system} were obtained through singular value
deconvolution~\cite{GSL} which was found to give satisfactory results
without requiring regularization via the projected Landweber
iteration. As previously, only even values of $l$ were included in
\eqref{eq:pbasex_system}.

\begin{figure}
  \includegraphics{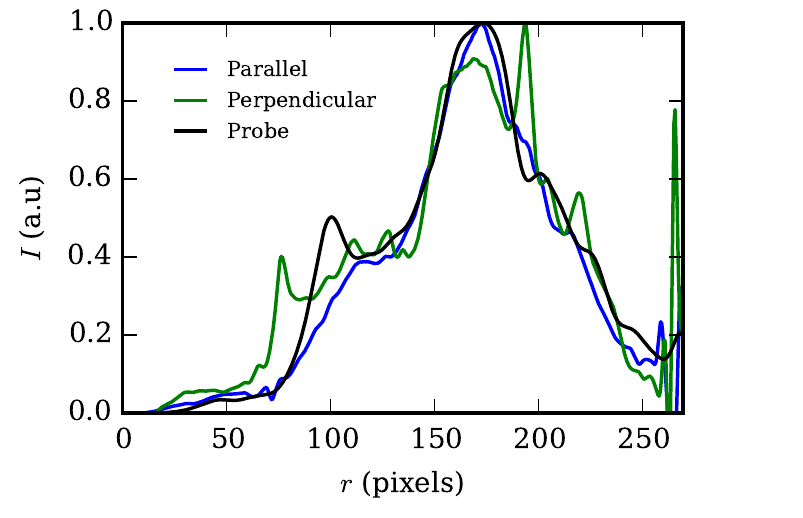}
  \caption{Radial spectra obtained from the pBasex inversion of the
    aligned molecule data (\figref{fig:aligned_vmi_2}) for the
    parallel (blue) and perpendicular (green) probing geometries. The
    radial spectrum obtained from pBasex inversion of the probe-alone
    data for randomly oriented molecules (\figref{fig:probe2}) is also
    shown (black). All spectra shown are normalized to a maximum value
    of 1.}
  \label{fig:DIB_rspec_2}
\end{figure}

The radial spectra obtained from the deconvolution calculated
according to \eqref{eq:radial_spec} shown in \figref{fig:DIB_rspec_2}
clearly show that for the perpendicular probe geometry there is a
reduced relative contribution from the outer CEI channel,
corresponding to explosion of the triply charged parent ion, compared
to the doubly charged parent ion CEI channel. In addition, other
features are observed in the perpendicular geometry radial spectra
suggesting that the relative contributions from different
fragmentation pathways are dependent on molecular
orientation.~\cite{Xie2014} It is interesting to note that these
details would not be apparent from the raw VMI data before
deconvolution/inversion.

\begin{figure}
  \includegraphics{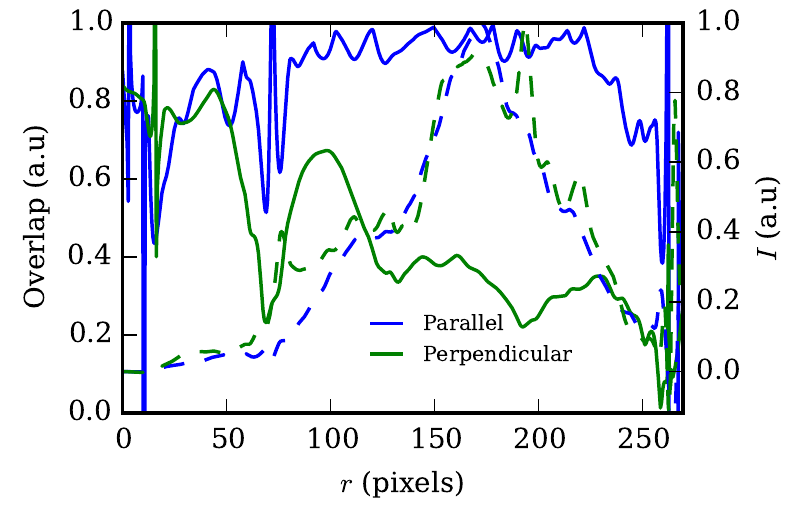}
  \caption{Radial dependence of the angular overlap factor
    corresponding to the pBasex inversion of the images shown in
    \figref{fig:aligned_vmi_2} (solid lines) for the parallel and
    perpendicular probing geometries. The corresponding radial spectra
    are also shown (dashed lines).}
  \label{fig:DIB_overlap2}
\end{figure}

In \figref{fig:DIB_overlap2} we show the radial dependence of the
angular overlap integral (\eqref{eq:overlap} evaluated as described in
\apxref{apx:overlap}). This plot shows that the perpendicular probing
geometry has a much lower angular overlap integral than the parallel
probe geometry, and also the circularly polarized probe described in
\subsecref{sec:circ_probe}(see \figref{fig:DIB_overlap}). This is due to
the fact that for this molecule the ionization probability for these
CEI channels is greatest when the probe laser polarization lies along
the I-I molecular axis. However, the ionization probability does not
drop to zero when the probe laser is perpendicular to the I-I
axis. One advantage of the perpendicular probe geometry is that, for
molecules with their I-I axes lying in the $yz$-plane, there is
uniform ionization probability with respect to molecular rotation
about the $x$ axis -- as such this provides a good measurement of the
degree of localization of the I-I molecular axes towards the $z$-axis
for those in-plane molecules. It is interesting to note that for the
perpendicular probe geometry, the inner (doubly charged parent)
channel has a higher overlap integral than the outer (triply charged
parent) channel, showing that the inner channel's dependence on
molecular orientation is weaker than that for the outer channel.

\begin{figure}
  \includegraphics{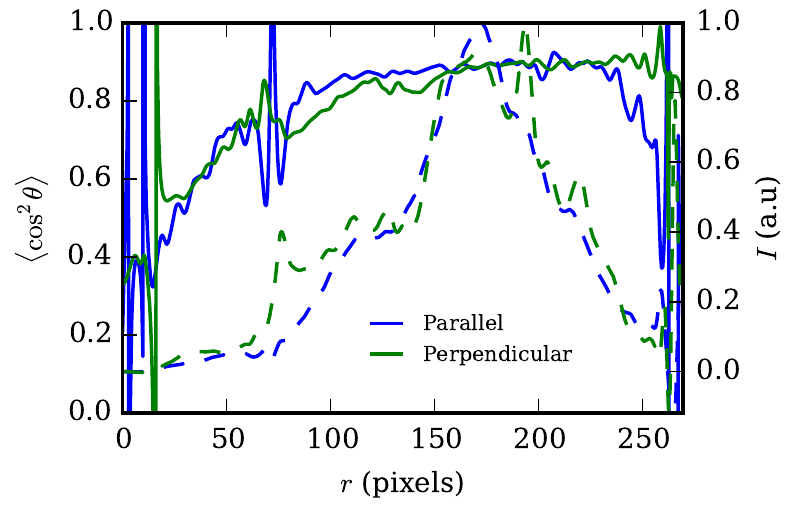}
  \caption{Radial dependence of the $\langle\cos^2\theta\rangle$
    expectation values obtained from the pBasex inversion of the data
    shown in \figref{fig:aligned_vmi_2} for the parallel (blue) and
    perpendicular (green) probe geometries. The radial spectra are
    also shown (dashed lines).}
  \label{fig:DIB_cos2_2}
\end{figure}

\figref{fig:DIB_cos2_2} shows the radial dependence of the
$\langle\cos^2\theta\rangle$ expectation value for the parallel and
perpendicular probe geometries calculated according to
\eqref{eq:cos2_expval}. For the outer channel the retrieved
$\langle\cos^2\theta\rangle$ is ca.~0.90 for both the parallel and the
perpendicular probe geometry. It might have been expected that the
higher angular overlap factor for the parallel probe geometry would
lead to a higher retrieved value of $\langle\cos^2\theta\rangle$ value
compared to that for the perpendicular probe geometry. On the other
hand in the parallel geometry the best aligned molecules have their
I-I axis close to the polarization axis of the probe pulse and as such
they have an increased probability of being ionized to higher charge
states due to enhanced ionization.~\cite{Seideman1995, Ellert1999,
  Hansen2012, Kjeldsen2005, Litvinyuk2003, Pavicic2007} These higher
charged molecular ions could fragment into I$^{n+}$ ions with $n>1$
rather than into I$^+$, i.e.~the best aligned molecules would not lead
to signal in the I$^+$ ion images and would therefore lead to a
reduced value of $\langle\cos^2\theta\rangle$ when determined from the
I$^+$ signal. The almost identical $\langle\cos^2\theta\rangle$ value
observed for the parallel and perpendicular geometries indicates that
neither the detection overlap factor nor enhanced ionization prevents
a reliable measurement of the degree of alignment for any of the probe
geometries using the algorithm presented here. In addition, it is
clear that although the relative weightings of different fragmentation
channels depends on the molecular orientation (as seen from the radial
spectra, \figref{fig:DIB_rspec_2}),~\cite{Xie2014} this is correctly
accounted for in the retrival algorithm presented, as evidenced by the
consistent $\langle\cos^2\theta\rangle$ values for the two probe
geometries. For the inner channel, the retrieved
$\langle\cos^2\theta\rangle$ value is lower for the perpendicular
probe geometry compared to the parallel probe geometry. Since both
probe geometries are sampling an identical molecular axis
distribution, the retrieved $\langle\cos^2\theta\rangle$ value should
be the same in both cases, as is observed for the outer channel. As
with the slightly reduced degree of alignment observed for the inner
channel when probing with the circularly polarized probe in
\subsecref{sec:circ_probe}, we attribute this difference as arising
due to non-axial recoil geometries being active for the inner
channel.~\cite{ChristensenPre}

\begin{figure*}
  \centering
  \includegraphics{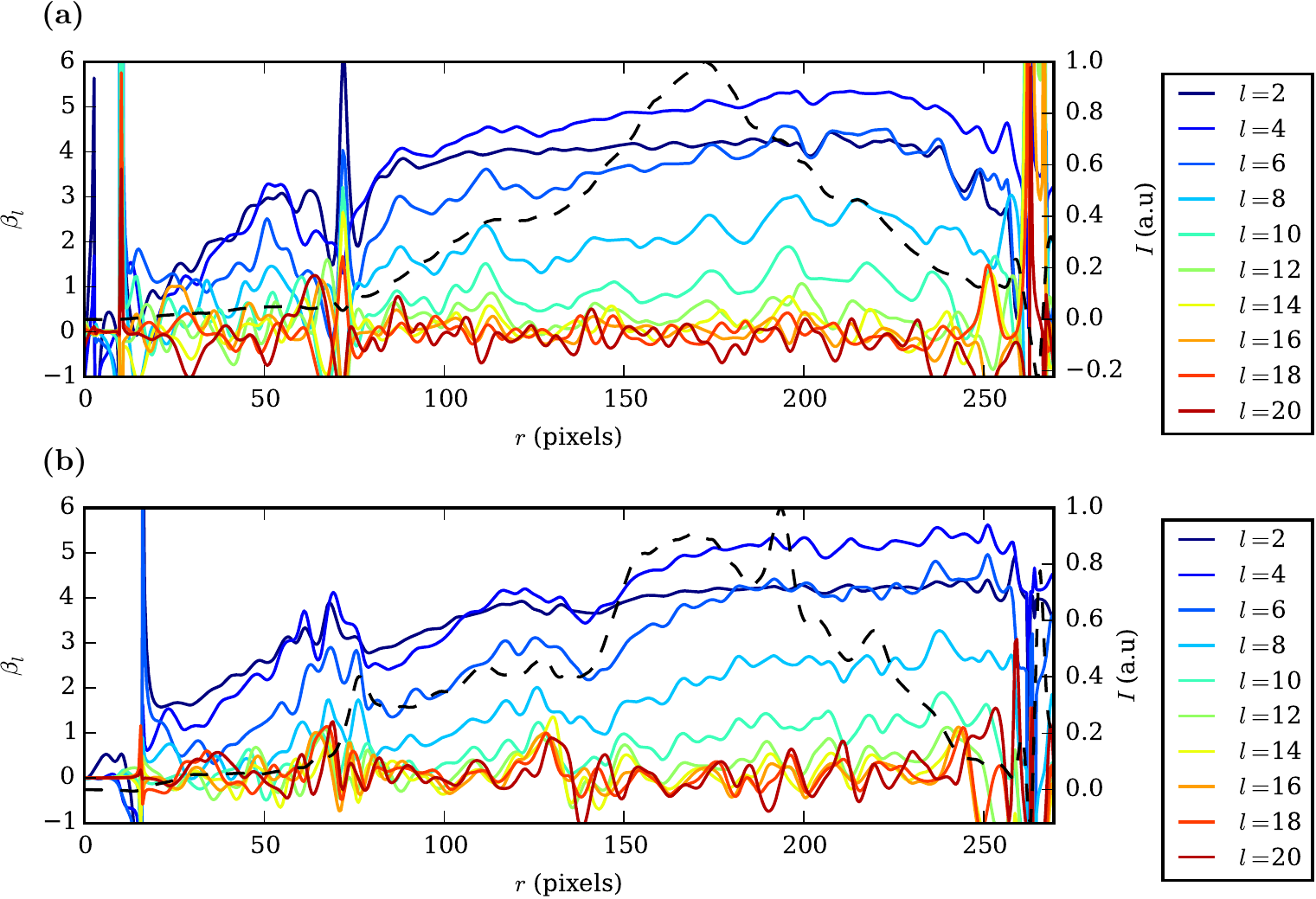}
  \caption{Radial dependence of the $\beta_l(r)$ angular parameters
    obtained from the deconvoluted pBasex inversion of the images
    shown in \figref{fig:aligned_vmi_2} (solid lines). (a) parallel
    probe polarization geometry (b) perpendicular probe polarization
    geometry.  The corresponding radial spectra are also shown (dashed
    lines).}
  \label{fig:DIB_beta2}
\end{figure*}

In \figref{fig:DIB_beta2} we show the $\beta_l(r)$ parameters for both
the parallel and perpendicular probe geometries calculated according
to \eqref{eq:beta_spec}. In both cases these coefficients show there
is negligible contribution from Legendre polynomials beyond 12 for the
alignment laser intensity used. The $\beta_l(r)$ coefficients for the
two probe geometries agree well for the outer CEI channel, but there
is a decrease in the $\beta_l(r)$ coefficients for the perpendicular
geometry similar to that seen with the $\langle\cos^2\theta\rangle$
expectation value.

\section{Conclusion}
We have proposed a method for deconvoluting a non-uniform detection
function from velocity map imaging experiments provided the detection
function is measurable independently. Experimentally we demonstrated
this technique by recovering the axis distribution of 1D aligned pDIB
molecules using laser-induced Coulomb explosion imaging. A major
advantage of the technique is that it allows a transferable and
complete characterization of the axis distribution of aligned
molecules.  In particular, $\langle\cos^2\theta\rangle$ can be
determined. This represents a measure of the true degree of alignment
rather than the usual $\langle\cos^2\Theta_{2D}\rangle$ value,
determined directly from 2D ion images, which is strongly biased by
the orientational dependence of the probe process. Furthermore, the
method also provides higher moments of the axis distribution and as
such a complete characterization alignment of the molecules is
possible.

\section{Acknowledgements}
We are grateful to Varun Suresh Makhija for helpful discussions on
this manuscript.

\appendix
\section{Abel inversion in spherical polar coordinates}
\label{apx:polar_abel}
For a specific value of $r$ (proportional to the particle velocity),
the relationship between the 3D distribution of detected particles,
$f(r, \theta, \phi) = g(r, \theta)D(r, \theta, \phi)$, and the
observed 2D projected image $F(R, \Theta)$ is
\begin{equation}
  \label{eq:Ff}
  F(R, \Theta;r)S_{R\Theta}=f(r, \theta, \phi)S_{\theta\phi}
\end{equation}
where $S_{R\Theta}$ is the elementary surface on the detector, and
$S_{\theta\phi}$ is the elementary surface on the sphere of radius
$r$,
\begin{align}
  S_{R\Theta} &= R\dee R\dee\Theta,\\
  \label{eq:Sthetaphi}
  S_{\theta\phi} &= r^2\sin\theta\dee\theta\dee\phi.
\end{align}
$S_{R\Theta}$ may be re-written as
\begin{equation}
  \label{eq:SRTheta}
  S_{R\Theta}=R|\mathbf{J}|\dee\theta\dee\phi,
\end{equation}
where the determinant of the Jacobian $\mathbf{J}$ is given by
\begin{equation}
  \label{eq:Jacobian}
  |\mathbf{J}| = 
  \left|
    \pd{R}{\theta}\pd{\Theta}{\phi}
    -
    \pd{\Theta}{\theta}\pd{R}{\phi}
  \right|.
\end{equation}
Substituting \eqref{eq:Sthetaphi}, \eqref{eq:SRTheta} and
\eqref{eq:Jacobian} into \eqref{eq:Ff} we can write
\begin{equation}
  \label{eq:FRTheta}
  F(R, \Theta;r)=
  \frac{f(r,\theta)r^2\sin\theta}{R|\mathbf{J}|}.
\end{equation}
Noting that
\begin{subequations}
  \begin{align}
    R&=r\sqrt{\cos^2\theta+\sin^2\theta\sin^2\phi},\\
    \Theta&=\arctan(\sin\phi\tan\theta),
  \end{align}
\end{subequations}
we can evaluate the Jacobian in \eqref{eq:Jacobian} as
\begin{equation}
  \label{eq:Jacobian2}
  |\mathbf{J}| = \left|\frac{2 r \sin ^2(\theta ) \cos (\phi )}
    {\sqrt{2 \cos (2 \theta ) \cos ^2(\phi )-\cos (2 \phi )+3}}\right|
  .
\end{equation}
Using the relationships
\begin{align}
  \label{eq:phi}
  \phi &= \arcsin \left(\frac{\tan\Theta}{\tan\theta}\right), \\
  \Theta &= \arccos \left(\frac{r\cos\theta}{R}\right),
\end{align}
we can subsitute \eqref{eq:Jacobian2} into \eqref{eq:FRTheta} to
obtain
\begin{equation}
  \label{eq:FRTheta2}
  F(R, \Theta; r)=\frac{rf(r, \theta, \phi)}{\sqrt{r^2-R^2}},
\end{equation}
with $\phi$ given by \eqref{eq:phi} and $\theta$ given by
\begin{equation}
  \label{eq:theta}
  \theta=\arccos\left(\frac{R\cos\Theta}{r}\right).
\end{equation}
Since in general we have more than a single kinetic energy present in
the 3D distribution, we have to integrate over $r\ge R$ in order to
calculate the projection intensity at $(R,\Theta)$:
\begin{equation}
  F(R, \Theta)=\int_{R}^{\infty}
  \frac{rf(r, \theta, \phi)}{\sqrt{r^2-R^2}}
  \dee r.
\end{equation}

\section{Rotation of the the detection function from the DF to the AF}
\label{apx:detfn_rotn}
The detection function in the DF (\eqref{eq:detfn_DF}) may be
re-written as an expansion in spherical harmonics as
\begin{equation}
  \label{eq:detfnshm}
  D(\theta', \phi'; r)=
  \frac{1}{\sqrt{4\pi}}
  \sum_{l=0}^{l_\mathrm{max}}
  \sqrt{\frac{4\pi}{2l+1}}
  \beta'_l(r)
  Y_{l0}(\theta', \phi').
\end{equation}
The detection function in the AF is related to the detection function
in the DF through rotation by the Euler angles $(\alpha, \Omega,
\gamma)$.~\cite{Zare} The detection function in the AF can be written as
\begin{equation}
  \begin{split}
    \label{eq:detfnrotaf}
    D(\theta, \phi; r)&=
    \frac{1}{\sqrt{4\pi}}
    \sum_{l=0}^{l_\mathrm{max}}
    \sqrt{\frac{4\pi}{2l+1}}
    \beta'_l(r)
    \\&\times
    \sum_{m=-l}^{m=l}
    D^l_{m0}(\alpha, \Omega, \gamma)
    Y_{lm}(\theta, \phi),
  \end{split}
\end{equation}
where $D^l_{mm'}(\alpha, \Omega, \gamma)$ are the Wigner rotation
matrices. Expressing the rotation matrix $D^l_{m0}(\alpha, \Omega,
\gamma)$ in terms of a spherical harmonic yields
\begin{equation}
  \begin{split}
    \label{eq:detfn_rotn_working}
    D(\theta, \phi; r)&=
    \frac{1}{\sqrt{4\pi}}
    \sum_{l=0}^{l_\mathrm{max}}
    \frac{4\pi}{2l+1}
    \beta'_l(r)
    \\&\times
    \sum_{m=-l}^{m=l}
    Y^\ast_{lm}(\Omega, \alpha)
    Y_{lm}(\theta, \phi),
  \end{split}
\end{equation}
The product of two spherical harmonics can be contracted by the
spherical harmonic addition theorem,\cite{Zare}
\begin{equation}
  \label{eq:shat}
  \frac{4\pi}{2l+1}
  \sum_{m=-l}^l
  Y^\ast_{lm}(\Omega,\alpha)
  Y_{lm}(\theta,\phi)
  =
  P_l(\cos\Delta),
\end{equation}
where $\Delta$ is given by \eqref{eq:detfn_loc}. Substitution of
\eqref{eq:shat} into \eqref{eq:detfn_rotn_working} gives
\eqref{eq:detfn_af}

\section{Detection function integrated over $\phi$}
\label{apx:detfnint}
In order to visualize how the detection function samples the axis
distribution, it is helpful to calculate the detection function in the
AF \eqref{eq:detfnrotaf} integrated over $\phi$. Noting that
\begin{equation}
  \int_0^{2\pi}
  Y_{lm}(\theta,\phi)
  \dee\phi
  =
  2\pi
  \sqrt{\frac{2l+1}{4\pi}}
  P_l(\cos\theta)\delta_{m0},
\end{equation}
and
\begin{equation}
  \label{eq:drotpl}
  D^l_{00}(\alpha, \Omega, \gamma)=
  P_l(\cos\Omega),
\end{equation}
we can evaluate the axially integrated detection function as
\begin{equation}
  \begin{split}
    D(\theta; r)&=
    \int_{0}^{2\pi}
    D(\theta, \phi; r)\dee\phi
    \\&=
    \sqrt{\pi}
    \sum_{l=0}^{l_\mathrm{max}}  
    \beta_l'(r)
    P_l(\cos\Omega)
    P_l(\cos\theta).
  \end{split}
\end{equation}

\section{Overlap function evaluation}
\label{apx:overlap}
The angular distribution of molecular axes in \eqref{eq:angdist} can
be re-expressed in terms of spherical harmonics as
\begin{equation}
  \label{eq:angdistshm}
  f(\theta;r)=\frac{1}{\sqrt{4\pi}}\sum_l\beta_l
  \sqrt{\frac{4\pi}{2l+1}}Y_{l0}(\theta,\phi).
\end{equation}
Substituting \eqref{eq:detfnrotaf} and \eqref{eq:angdistshm} into
\eqref{eq:overlap} gives
\begin{equation}
  \begin{split}
    O(r)&=\frac{1}{D_\mathrm{max}(r)}
    \sum_{l=0}^{l_\mathrm{max}}
    \sqrt{\frac{1}{2l+1}}
    \beta_l(r)
    \\&\times
    \sum_{l'=0}^{l'_\mathrm{max}}
    \sqrt{\frac{1}{2l'+1}}
    \beta'_{l'}(r)
    \sum_{m'=-l'}^{m'=l'}
    D^{l'}_{m'0}(\alpha, \Omega, \gamma)
    \\&\times
    \int^{2\pi}_0
    \int^{\pi}_0
    Y_{l0}(\theta,\phi)
    Y_{l'm'}(\theta, \phi)
    \sin\theta\dee\theta\dee\phi,
  \end{split}
\end{equation}
Evaluating the integral using the orthogonality of spherical
harmonics~\cite{Zare} and \eqref{eq:drotpl} gives
\begin{equation}
  \label{eq:overlap3}
  O(r)=\frac{1}{D_\mathrm{max}(r)}
  \sum_l
  \frac{1}{2l+1}
  \beta_l(r)\beta'_l(r)P_l(\cos\Omega).
\end{equation}

\bibliography{refs}

\end{document}